\documentclass[twocolumn,showpacs,preprintnumbers,amsmath,amssymb]{revtex4}
\usepackage{graphicx}
\usepackage{dcolumn}
\usepackage{bm}

\def\ket#1{\lvert#1\rangle}
\def\bra#1{\langle #1\lvert}

\def\ketbra#1#2{\lvert#1\rangle\langle#2\lvert}
\def\vek#1{{\bf{#1}}}

\begin{document}


\title{Violation of a Bell-like inequality for spin-energy entanglement in neutron polarimetry}
\author{S. Sponar$^1$}
\author{J. Klepp$^{2}$}
\author{C. Zeiner$^1$}
\author{G. Badurek$^1$}
\author{Y. Hasegawa$^{1}$}
\affiliation{%
$^1$Atominstitut der \"{O}sterreichischen Universit\"{a}ten, 1020
 Vienna, Austria\\ $^2$Faculty of Physics, University of Vienna,
Boltzmanngasse 5, A-1090 Vienna, Austria  }

\date{\today}

\begin{abstract}
Violation of a Bell-like inequality for a spin-energy entangled neutron state has been demonstrated in a polarimetric experiment. The proposed
inequality, in Clauser-Horne-Shimony-Holt (CHSH) formalism, relies on correlations between the spin and energy degree of freedom in a
single-neutron system. The correlation function $S_{CHSH}$ is determined to be 2.333\,$\pm$\,0.002 $\not \leq2$, which violates the Bell-like
CHSH inequality by more than 166 standard deviations.
\end{abstract}


\maketitle

Einstein, Podolsky and Rosen (EPR) argued, based on the assumption of local realism, that quantum mechanics (QM) is not a complete theory
\cite{EPR35}. In 1951 Bohm reformulated the EPR argument for spin observables of two spatially separated entangled particles to illuminate the
essential features of the EPR paradox \cite{BOHM51}. Thereafter Bell introduced inequalities for certain correlations which hold for the
predictions of any local hidden variable theory (LHVT) applied \cite{Bell64}, but are violated by QM. From this, one can conclude that QM cannot
be reproduced by LHVTs. Five years later Clauser, Horne, Shimony and Holt (CHSH) reformulated Bell\char39{}s inequalities pertinent for the
first practical test of quantum non locality \cite{Clauser69}. Polarization measurements with correlated photon pairs
\cite{BookBertlmannZeilinger}, produced by atomic cascade \cite{Freedman72,Aspect82} and parametric down-conversion of lasers
\cite{Kwiat95,Weihs98,Gisin98}, demonstrated violation of the CHSH inequality. Up to date many physical systems
\cite{Rowe01,moehring04,sakai06,Matsukevich2008} have been examined, including neutrons in an interferometric experiment \cite{Hasegawa03Bell}.

Most EPR experiments test LHVTs, which are based on the assumptions of locality and realism. LHVTs are a subset of a more general class of
hidden-variable theories, namely the noncontextual hidden-variable theories (NCHVTs). Noncontextuality implies that the value of a dynamical
variable is determined and independent of the experimental context, i.e. of previous or simultaneous measurements of a commuting observable
\cite{Bell66,Mermin93}. Noncontextuality is a more stringent demand than locality because it requires mutual independence of the results for
commuting observables even if there is no spacelike separation \cite{Simon00}.

In the case of neutrons entanglement is not achieved between particles but between different degrees of freedom (DOF). Since the observables of
one Hilbert spaces (HS), describing a certain DOF, commute with observables of a different HS, the single neutron system is suitable for
studying NCHVTs with multiple DOF. Using neutron interferometry \cite{Rauch00Book}, single-particle entanglement between the spinor and the
spatial part of the neutron wavefunction \cite{Hasegawa03Bell}, as well as full tomographic state analysis \cite{hasegawa2007tomography}, have
already been accomplished. In addition, the contextual nature of quantum theory \cite{Hasegawa2006contextual,Bartosik2009} has been
demonstrated. In a recent experiment creation of a triply entangled single neutron state was achieved by developing a coherent-manipulation
method for the total energy (the sum of kinematic and potential energies) of a single neutron system \cite{Sponar07}.

Neutron polarimetry has several advantages compared to perfect crystal interferometry. It is insensitive to ambient mechanical and thermal
disturbances and therefore provides better phase stability. Efficiencies of the manipulations, including state splitting and recombination, are
considerably high (typically $>$98\,\%) resulting in a better contrast compared to interferometry. In addition, Single-crystal interferometers
accept neutrons only within an angular range of a few arcseconds, which leads to a significant decrease in intensity. Neutron polarimetry has
been used to demonstrate the noncommutation properties of the Pauli spin operator \cite{Hasegawa99NoncommutingSpinor}, and for geometric phase
measurements \cite{Badurek00SeperationGeoDyn,Klepp05MixedPancharatnamPhase,Klepp08}. Polarimetry is applicable to other quantum systems aside
from neutrons, for instance Ca \cite{Yanagimachi02} or $^3$He \cite{Kamiya03}. Further developments within the scope of atomic polarimetry are
anticipated.

This letter reports the first experimental confirmation for the violation of a Bell-like CHSH inequality in the field of neutron polarimetry.
The inequality is based on the statistical outcome of correlation measurements of the spin and energy DOF.
\begin{figure*}[t]
\begin{center}
\scalebox{0.45}{\includegraphics{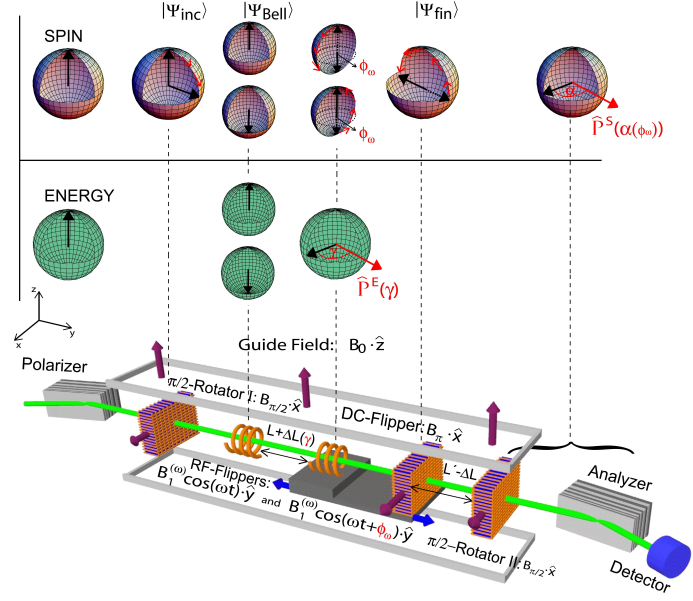}} \end{center}  
 \caption {Experimental apparatus for observation of quantum correlations between the spin and energy degree of freedom, expressed by a
 Bell-like CHSH inequality. The incident neutron beam is polarized by a supermirror polarizer. A DC-$\pi/2$ spin-turner creates a spin superposition followed by
an RF-flipper, preparing the entanglement of spin and energy. The position change of the translation stage (displacement of second RF-flipper
together with DC spin-flipper) and the phase difference between the two oscillating RF-fields adjust the parameters $\alpha$ and $\gamma$ for
the Bell measurement. Finally, the spin is projected back to the initial direction using a second DC-$\pi$/2 spin-turner for a spin polarization
analysis followed by a count rate detection. The Bloch-sphere description depicts the evolutions of each quantum state in spin and energy
subspaces. It includes measurement settings of $\alpha$ and $\gamma$, determining the projection operators used for joint measurement of spin
and energy. The effect of the DC-flipper on the translation stage is to suppress a change in the total Larmor precession angle (see main text).}
\label{fig:setup}
\end{figure*}

In our experiment the neutron\char39{}s wavefunction exhibits entanglement between the spinor and energy DOF \cite{Sponar07}, expressed as
\begin{equation}\label{eq:ExpectationValue}
\ket{\Psi_{\textrm{Bell}}}=\frac{1}{\sqrt{2}}\Big(\ket{E_0+\hbar\omega}\otimes \ket{\uparrow} + \ket{E_0-\hbar\omega}\otimes
\ket{\downarrow}\Big).
\end{equation}
Here $\ket{\uparrow}$ and $\ket{\downarrow}$ denote the neutron\char39{}s up and down spin eigenstates, referring to the chosen quantization
axis, describing a spin superposition prepared by a DC-$\pi$/2 spin-turner. $\ket{E_0+\hbar\omega}$ and $\ket{E_0-\hbar\omega}$ are the energy
eigenstates after interaction with a time-dependent magnetic field within a radio frequency (RF)-flipper, due to absorption or emission of
photons of energy $\hbar\omega$. $E_0$ is the initial total energy of the neutron and $\omega$ is the frequency of the oscillating magnetic
field.

As in common Bell experiments a joint measurement of two observables, i.e. spin and energy, is performed: First projection operators onto a spin
superposition state, specified by an angle parameter $\alpha$, are defined as
\begin{equation}
\hat P^{(S)}_{\pm}(\alpha)=\frac{1}{\sqrt2}\big(\ket{\uparrow}\pm e^{-i\alpha}\ket{\downarrow}\big)\big(\bra{\uparrow}\pm
e^{i\alpha}\bra{\downarrow}\big).
\end{equation}
Similarly projection operators onto an energy superposition state, using an angle parameter $\gamma$, are given by
\begin{eqnarray}\label{eq:expecttheory}
\hat P^{(E)}_{\pm}(\gamma)&=&\frac{1}{\sqrt2}\big(\ket{E_0+\hbar\omega}\pm e^{-i\gamma}\ket{E_0-\hbar\omega}\big)\\\nonumber
&\times&\big(\bra{E_0+\hbar\omega}\pm e^{i\gamma}\bra{E_0-\hbar\omega}\big).
\end{eqnarray}
The angle parameters $\alpha$ and $\gamma$ are the azimuthal angles on the Bloch spheres corresponding to the spin and energy DOF, respectively.
They are depicted in Fig.\,\ref{fig:setup}. Introducing the observables
\begin{equation}
\hat A^{(S)}(\alpha)=\hat P^{(S)}_{+}(\alpha)-\hat P^{(S)}_{-}(\alpha)
\end{equation}
and
\begin{equation}
\hat B^{(E)}(\gamma)=\hat P^{(E)}_{+}(\gamma)-\hat P^{(E)}_{-}(\gamma)
\end{equation}
one can define an expectation value for a joint measurement of spin and energy along the directions $\alpha$ and $\gamma$
\begin{eqnarray}\label{eq:expectValue}
 E(\alpha,\gamma)&=&\bra{\Psi(t)}\hat A^{(S)}\otimes \hat B^{(E)}\ket{\Psi(t)}=\cos(\alpha+\gamma).
\end{eqnarray}
For a Bell-like inequality in CHSH-formalism \cite{Clauser69} four expectation values as defined in Eq.(\ref{eq:expectValue}), with the
associated directions $\alpha_1, \alpha_2$ and $\gamma_1, \gamma_2$ for joint measurements of spin and energy, are required which yields
\begin{eqnarray}\label{eq:SBell}
&&S_{CHSH}(\alpha_1,\alpha_2,\gamma_1,\gamma_2)\\\nonumber &&=\vert
E(\alpha_1,\gamma_1)+E(\alpha_2,\gamma_1)-E(\alpha_1,\gamma_2)+E(\alpha_2,\gamma_2)\vert.
\end{eqnarray}
The boundary of Eq.(\ref{eq:SBell}) is given by the value 2 for any NCHVT, whereas QM predicts a maximal value $S_{CHSH}^{MAX}$=$2\sqrt2$ for
$\alpha_1=0$, $\alpha_2=\pi/2$, $\gamma_1=\pi/4$ and $\gamma_2=3\pi/4$.

The experiment was carried out at the tangential beam port of the 250 kW TRIGA research reactor of the Atomic Institute of the Austrian
Universities, Vienna. A schematic view of the experimental arrangement is shown in Fig.\,\ref{fig:setup}.

A neutron beam of mean wavelength $\lambda=$1.99 $\mbox{\AA}$, reflected from a pyrolytic graphite monochromator and propagating in the
$+\hat{\mathbf y}$-direction, is polarized along the $\hat{\mathbf z}$-direction using a bent Co–Ti supermirror array. The first DC-coil,
functioning as a $\pi/2$ spin-turn device, rotates the spin into the $\hat{\mathbf x}\hat{\mathbf y}$-plane. Thus a coherent superposition of
the two orthogonal spin eigenstates$\ket{\uparrow_z}$ and $\ket{\downarrow_z}$ in equal portions is created, yielding an incident state denoted
as
\begin{eqnarray}\label{eq:stateinc}
\ket{\Psi_{\textrm{inc}} }= \frac{1}{\sqrt{2}}\Big ( \ket{\uparrow_z}+\ket{\downarrow_z}\Big)\otimes \ket{E_0 }.
\end{eqnarray}

The entanglement between the spinor and energy DOF is created exploiting the mode of operation of a subsequent RF-flipper \cite{Sponar07} with
an oscillating field $B(t) = B^{(\omega)}_1 \cos(\omega t)\cdot\hat{\mathbf y}$. Fulfilling the resonance condition ($\omega=2\vert\mu\vert
B_0/\hbar$) for the oscillating field and the guide field, the $\hat{\mathbf z}$-component of the total magnetic field can be completely
suppressed within the rotating frame of the oscillating field. The effective field, perpendicular to the initial polarization, is adjusted to
$B^{(\omega)}_1=\pi\hbar/(2\tau\vert\mu\vert)$ initiating a spin-flip process. Here $\mu$ is the magnetic moment of the neutron and $\tau$ is
the time the neutron requires to traverse the RF-field region. Interacting with a time-dependent magnetic field, the total energy of the neutron
is no longer conserved due to absorption and emission of photons of energy $\hbar\omega$, depending on the spin state
\cite{Summhammer93MultiPhoton,Golub94}. The RF-flipper is operating at a frequency of $\omega/2\pi=32$\,kHz and accordingly the guide field is
tuned to $B_0\sim 1.1$\,mT. The entangled state vector can be represented as a Bell state
\begin{eqnarray}\label{eq:statebell}
\ket{\Psi_{\textrm{Bell}} }&&=\\\nonumber&& \frac{1}{\sqrt{2}}\Big ( \ket{E_0+\hbar\omega}\otimes \ket{\uparrow_z}+ \ket{E_0-\hbar\omega}\otimes
\ket{\downarrow_z}\Big),
\end{eqnarray}
which is illustrated using a Bloch sphere description in Fig\,\ref{fig:setup}.
\begin{figure}[b]
\begin{center}
\scalebox{0.4}{\includegraphics {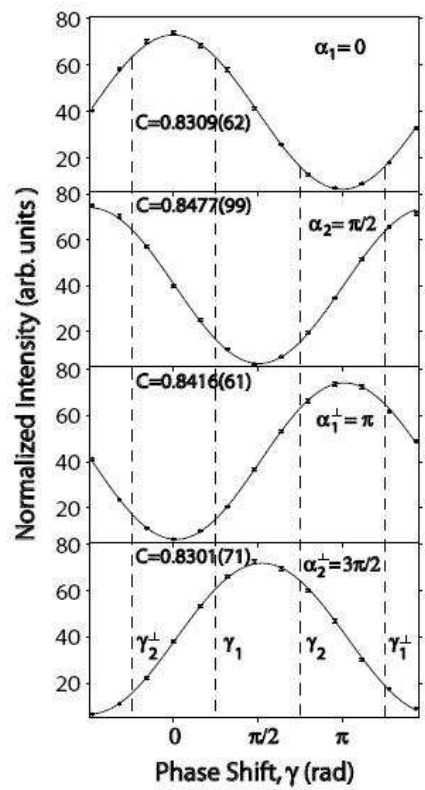}}\end{center} \caption {Typical interference oscillations, due to a variation of $\gamma$, for
$\alpha_1$ = 0, $\alpha_2=\pi/2$, ($\alpha_1^\bot=\pi$, $\alpha_2^\bot=3\pi/2$). One period corresponds to a displacement of 31.28\,$\pm$\,0.06
mm of the translation stage. The dashed lines mark the $\gamma$ values of $\gamma_1=\pi/4, \gamma_2=3\pi/4$ $ (\gamma_1^\bot=5\pi/4$,
$\gamma_2^\bot=7\pi/4=-\pi/4$), where a maximum violation of the Bell-like CHSH inequality is expected. The joint measurements of expectation
values exhibit $S_{CHSH}$ =2.333\,$\pm$\,0.002.}\label{fig:oscillations}
\end{figure}

\begin{center}
\begin{table*}\caption{\label{tab:table1}Results of the spin-energy correlation measurements.}
\begin{tabular}{ccc}
$\alpha_i\, (\alpha_i^\bot)$&$\gamma_j \,(\gamma_j^\bot)$&$E$($\alpha_i,\gamma_j$) \hspace{10mm}$(i,j=1,2)$\\ \hline
 $\alpha_1=0\,$($\pi$)&$\gamma_1=\pi/4$\,($5\pi/4$)&$E(\alpha_1,\gamma_1)$\,=\,0.594  $\pm$ 0.001  \\
 $\alpha_2=\pi/2$\,($3\pi/2$)&$\gamma_1=\pi/4$\,($5\pi/4$)&$E(\alpha_2,\gamma_1)$\,=\,0.575 $\pm$ 0.001\\
 $\alpha_1=0\,$($\pi$)&$\gamma_2=3\pi/4$\,($7\pi/4$)&$E(\alpha_1,\gamma_2)$\,=\,-0.571 $\pm$ 0.001 \\
 $\alpha_2=\pi/2$\,($3\pi/2)$&$\gamma_2=3\pi/4$\,($7\pi/4$)&$E(\alpha_2,\gamma_2)$\,=\,0.593 $\pm$ 0.001\\\hline\hline
 &$\hspace {11mm}S_{CHSH}$= 2.333 $\pm$ 0.002
\end{tabular}
\end{table*}
\end{center}
The second RF-flipper and an auxiliary DC-flipper are mounted on a (single) translation stage. The function of the DC-flipper is explained
below. By choosing the same frequency for the two RF-flippers the energy difference between the two spin components is compensated. The
oscillating field of the second RF-flipper is given by $B(t) = B^{(\omega)}_1 \cos(\omega t+\phi_\omega)\cdot\hat{\mathbf y}$. This procedure is
described by the action of the projection operator for the energy recombination $\hat
P^{(E)}=\ket{E_0}\big(\bra{E_0+\hbar\omega}+(\bra{E_0-\hbar\omega}\big)$. Applied to Eq.(\ref{eq:statebell}), and considering the second
RF-flipper(energy recombination) and the DC-flipper, this operator yields the final state
\begin{eqnarray}\label{eq:statef}
\ket{\Psi_{\textrm{fin}} }= \frac{1}{\sqrt{2}}\Big ( e ^{-i\phi_\omega}\ket{\uparrow_z}+ e ^{i\omega T}e^{i\phi_\omega}
\ket{\downarrow_z}\Big)\otimes \ket{E_0 }.
\end{eqnarray}
Here $\omega\cdot T\equiv\gamma$ is the phase acquired in energy subspace, where $T$ is the propagation time for the distance $L+\Delta L$
between the two RF-flippers and $\phi_\omega$ is the tuneable phase of the oscillating field of the second RF-flipper.

The stationary guide field $B_0\cdot\hat{\mathbf z}$ induces an additional phase due to Larmor precession within the guide field region. However
this phase contribution remains constant during the experiment and can therefore be adjusted by finding the zero-position of $\gamma$ scans
(displacement of translation stage). The corresponding polarization vector is given by
\begin{equation}\label{eq:k}
\vek{p}_{\textrm{fin}} =\Big(\cos\big(\gamma+\alpha\big),\sin\big(\gamma+\alpha\big),0\Big),
\end{equation}
with the spin phase $\alpha\equiv2\phi_\omega$, originating from the phase of the oscillating field of the second RF-flipper (a detailed
description of the spin phase acquired due to a spinor evolutions from the north to the south pole of the Bloch sphere, and back, is given in
\cite{Badurek00SeperationGeoDyn}). The phases $\alpha$ and $\gamma$ are associated with the measurement directions on the equatorial plane of
the Bloch spheres, required for joint measurements of spin and energy.

Compensation of the Larmor phase is accomplished by the auxiliary DC-flipper placed subsequently to the second RF-flipper. No additional phase
shift, induced by Larmor precession, resulting from the change of $\Delta L$ is observed. Phase contributions with the same sign occur in the
regions $L+\Delta L$ (between first and second RF-flippers) and $L^\prime-\Delta L$ (between DC-flipper and second DC-$\pi/2$ spin-turner)
compensating each other. Therefore the total Larmor rotation angle $\big(\propto \omega_{\textrm{L}}(L+L^\prime)\big)$ remains constant although
the position of the translator is altered (see \cite{Sponar08} for details of this procedure). Consequently only $\gamma$, the phase of the
energy subspace, is affected by a displacement of the translation stage.

The second DC-$\pi/2$ spin-turner reverses the action of the first one by a -$\pi/2$ spin-rotation around the $+\hat{\mathbf x}$ axis for a
forthcoming polarization analysis along the +$\hat{\mathbf z}$ direction by the second supermirror. This is expressed by applying a projection
operator for the spin $\hat P^{(S)}=\ketbra{\uparrow_z}{\uparrow_z}$. Finally the stationary intensity oscillations are given by
\begin{equation}\label{eq:Intensity}
N(\alpha,\gamma) =\frac{1}{2}\Big( 1+C\cos\big(\alpha+\gamma)\Big),
\end{equation}

where $C$ is the contrast, which is 100\,\%(C=1) under ideal circumstances. In our setup $C$ was experimentally determined as
83.8\,$\pm$\,0.4\%, which is depicted in Fig\, \ref{fig:oscillations}. The physical reasons for the loss in contrast are explained later. Thus
the expectation value, defined in Eq.(\ref{eq:expectValue}), can be rewritten using the normalized count rates obtained with the measurement
settings of $\alpha$ and $\gamma$ denoted as
\begin{equation}\label{eq:expectExperimentCounts}
\!\!\!\!\!\!E(\alpha,\gamma)=\frac{N(\alpha,\gamma)+N(\alpha^\bot,\gamma^\bot)-N(\alpha,\gamma^\bot)-N(\alpha^\bot,\gamma)}
{N(\alpha,\gamma)+N(\alpha^\bot,\gamma^\bot)+N(\alpha,\gamma^\bot)+N(\alpha^\bot,\gamma)},
\end{equation}
with $\alpha^\bot=\alpha+\pi$ and $\gamma^\bot=\gamma+\pi$. Therefore, from the contrast C=0.838 a value of 2.37($0.838\cdot2\sqrt{2}\sim2.37$)
is expected for $S_{CHSH}$ for $\alpha_1=0$, $\alpha_2=\pi/2$, $\gamma_1=\pi/4$ and $\gamma_2=3\pi/4$.

Typical oscillations, observed when the position of the translation stage (second RF-flipper) is varied ($\gamma$-scans), are plotted in Fig\,
\ref{fig:oscillations} for different settings of $\alpha$. One period corresponds to a displacement of the translator stage of
31.28\,$\pm$\,0.06 mm. The $\gamma$-scan for $\alpha_1=0$ was used to determine the position of the translation stage corresponding to the
values $\gamma_1=\pi/4, \gamma_2=3\pi/4$ $ (\gamma_1^\bot=5\pi/4$, $\gamma_2^\bot=7\pi/4=-\pi/4$) which are, together with the spin phase
settings  $\alpha_1=0$, $\alpha_2=\pi/2$ ($\alpha_1^\bot=\pi$,$\alpha_2^\bot=3\pi/2$), required for determining the $S$-value for a maximal
violation of the Bell-like CHSH inequality.

The actual Bell measurement consists of successive count rate measurements using appropriate settings of the phase of the oscillating field of
the second RF-flipper (tuning the spin phase $\alpha$) and position of the second RF-flipper mounted on the translation stage (tuning the energy
phase $\gamma$). The four expectation values $E(\alpha_i,\gamma_j)$ ($i,j=1,2$), for joint measurement of spin and energy DOF, are determined
from the associated count rates, using Eq.(\ref{eq:expectExperimentCounts}). They are listed in Table \ref{tab:table1}. After three complete
measurement sets (to reduce statistical errors) a final value $S_{CHSH}$=2.333\,$\pm$\,0.002 was determined which is notedly above the value of
2, predicted by NCHVTs and close to 2.37, a value derived by taking a contrast of 83.8\,\% of the interferograms into account.

A slight deviation of the measured $S_{CHSH}$ value of 2.333\,$\pm$\,0.002 from the expected value of 2.37 can be explained by inhomogeneities
of the guide field $B_0$, which results in fluctuations of the energy phase $\gamma$, as well as imperfections of the spin phase manipulation.
(Varying the phase $\phi_\omega$ of the oscillating field of the second rf-flipper e.g. by $\pi/4$ should theoretically yield a phase shift of
the intensity modulations of $\pi/2$, as predicted by Eq.(\ref{eq:Intensity}), whereas in practice we measured additional shifts around one and
two degrees from the desired settings.) Due to the deviation of the measured $S_{CHSH}$ value from the predicted value it is useful to introduce
an error estimation of the calculated value of 2.37, which consists of three parts: The error of the contrast measurement (83.8\,$\pm$\,0.4\%),
inhomogeneities of the guide field $B_0$, and imperfections of the spin phase manipulation. These contributions lead to a final error of the
calculated value estimated by $\sim$\,0.036. This error (2.37\,$\pm$\,0.036) is one magnitude larger compared to the error of the Bell
measurement (2.333\,$\pm$\,0.002). The former reflects all systematic imperfections of the setup, whereas the latter is solely a statistic error
derived from the count rates, which are very high (up to 32000 cnts per point) due to a long measurement time.

The average contrast of C\,=\,83.8\,\%, of the observed intensity oscillations, exceeds the minimum visibility $C_{{crit}}=
70.7\,\%\,(\sqrt2/2)$ necessary to exhibit a violation of the CHSH inequality. The maximal $S$-value that can be achieved experimentally is
proportional to the contrast and in our case given by $ C\cdot S_{CHSH}^{MAX}=2.37>2$ since  $C>C_{{crit}}$. The main reason for the rather low
contrast with respect to the high flip efficiencies of the DC and RF-flippers is a broad momentum distribution of $\sim 2$\,\% induced by the
mosaic structure of our monochromator crystal. This corresponds to a broad distribution of the propagation-time, and therefore the distribution
of the spin-rotation angle (in the $\hat{\mathbf x}\hat{\mathbf y}$-plane) is widened after each full rotation of the polarization vector, which
results a noticeable decrease in contrast.

In our setup neutrons are detected by a BF$_3$ detector with an inherent efficiency of $>$99\,\%, which is larger than the well known threshold
efficiency $\eta_{{crit}}=2(\sqrt{2}-1)\sim\,0.83$ required to close the detection loophole with maximally entangled states
\cite{Garg87,Larsson98}.

To summarize, we have demonstrated a violation of a Bell-like CHSH inequality in the field of polarimetry with massive spin-$\frac{1}{2}$
particles i.e. neutrons. The measured correlations between the neutron\char39{}s spinor and energy DOF contribute to a result $S_{CHSH}$=
2.333\,$\pm$\,0.002, which exceeds the Bell limit, demarcating NCHVTs, by more than 166 standard deviations. The technique established here will
be utilized in forthcoming neutron optical experiments for preparation of multi-entanglement in neutron polarimetry and interferometry.

This work has been partly supported by the Austrian Science Foundation, FWF (P21193-N20).

\end{document}